\documentclass[%
 reprint,
 amsmath,amssymb,
 aps,
]{revtex4-1}

\usepackage{graphicx}
\usepackage{dcolumn}
\usepackage{bm}
\begin{document}

\title{Emergence of Flat-Band Magnetism and Half-Metallicity in Twisted Bilayer Graphene}

\author{Alejandro Lopez-Bezanilla}
\email[]{alejandrolb@gmail.com}
\affiliation{ Theoretical Division, Los Alamos National Laboratory, Los Alamos, New Mexico 87545, United States}

\date{\today }

\begin{abstract}
Evidence of flat-band magnetism and half-metallicity in compressed twisted bilayer graphene is provided with first-principles calculations. 
We show that dynamic band-structure engineering in twisted bilayer graphene is possible by controlling the chemical composition with extrinsic doping, the interlayer coupling strength with pressure, and the magnetic ordering with external electric field.
By varying the rotational order and reducing the interlayer separation an unbalanced distribution of charge density resulting in the spontaneous apparition of localized magnetic moments without disrupting the structural integrity of the bilayer. Weak exchange correlation between magnetic moments is estimated in large unit cells. 
External electric field switches the local magnetic ordering from ferromagnetic to anti-ferromagnetic. 
Substitutional doping shifts the chemical potential of one spin distribution and leads to half-metallicity.  Flakes of compressed twisted bilayer graphene exhibit spontaneous magnetization, demonstrating that correlation between magnetic moments is not a necessary condition for their formation.    
\end{abstract}

\keywords{Twisted bilayer graphene, magnetism, half-metallicity, doping}
\maketitle

\section{\label{introduction}Introduction}

Progress in assembling multilayered two-dimensional (2D) materials has come with a rich array of physics arising from the van der Waals (vdW) forces that hold together parallel layers\cite{Ajayan2D}. Owing to the geometrical degrees of freedom allowed by the vdW interactions, flat structures can easily change the layers' mismatch by rotating the planes and to vary the interlayer distance by applying compression. As a result, compounds with physical properties that are fundamentally distinct from those of the constituent layers can be created. 

Special attention deserves twisted bilayer graphene (tBLG), an extended crystalline planar structure composed of two vdW bonded graphene sheets where one sheet is twisted with respect to the other. At specific rotational angles, known as {\it magic angles}, the Moir{\'e} patterns that emerge as a result of the offset in the relative orientation of the sheets create an atomic-scale periodic potential able to modify the linear dispersion of graphene\cite{Koshino}. The flat bands created at the twist angle of 1.05$^\circ$ allow tBLG to exhibit an unconventional superconducting phase not observed in monolayer or AB stacked bilayer graphene \cite{Jarillo1,Jarillo2,Ivar,Stauber,Calderon,Uchoa}.
MIT's Jarillo-Herrero's group reported an insulating state when injected electrons reside in a Moir{\'e} unit cell, relating this behavior to an example of Mott physics\cite{Jarillo1,Jarillo2}. Padhi {\it et al.} ascribed this physical effect to a Wigner crystallization\cite{Padhi2018}.
Strongly correlated states in compressed tBLG far from the flat-band angle were measured by Yankowitz and co-workers. \cite{Yankowitz,Yankowitzeaav1910}, pointing out to an avenue to creating partially filled flat bands at higher twist angles.  In all cases, hints of strong-coupling behavior appears at twist angles less than 2$^\circ$. The possibility of creating magnetic structures in compressed tBLG with a larger twist angle was suggested by Yndur{\'a}in\cite{Felix}. 
The existence of intrinsic magnetism in graphitic layers is still debated and generally related to the presence of topological defects such as zigzag edges, grain boundaries, vacancies or adatoms\cite{Ugeda,Asenjo}, which break the sublattice symmetry and create a magnetic moment of 1$\mu_B$ per isolated p$_z$ orbital removed\cite{Lieb}. Hitherto, the possibility of inducing magnetism in 2D bilayer graphene without disrupting the hexagonal network has been suggested and remains poorly characterized with parameter-free approaches. Parametrized tight-binding calculations within the Hubbard model of non-relaxed small-twist angle tBLG suggested that anti-ferromagnetism and spiral ferromagnetism may arise\cite{Arraga}. However, the uncertainty of the on-site repulsion term that eventually leads to magnetism in model Hamiltonians often yields ambiguous results. It is therefore highly desirable to clarify the conditions for observing magnetism in bilayer graphitic networks within an {\it ab initio} approach to treat realistic systems in their detailed complexity using a minimal set of tuning physical parameters.

In this paper we study the physical properties of tBLG with twist angles in the range of 5.05$^\circ$ to 2.0$^\circ$. Theoretical evidence of the emergence of magnetism under pressure and doping is provided, and the tunability of the bilayers' magnetic properties considered. We find that the transient distortions in the distribution of charge density between vdW bonded layers are overwhelmed by the enhanced hybridization of AA stacked C atoms under compressing strain, which leads to a sudden flattening of the dispersive bands and to an unbalanced occupation of the localized states. The spontaneous emergence of localized magnetic moments is analyzed by rotating systematically one of the layers, observing that magnetism is induced only within a small range of twist angles. The magnetic arrangement of the localized states is determined by subtle differences in energy with a competition between ferro- and anti-ferromagnetic ordering between moments at AA stacking zones.

\begin{figure*}[htp]
 \centering
   \includegraphics[width=0.95 \textwidth]{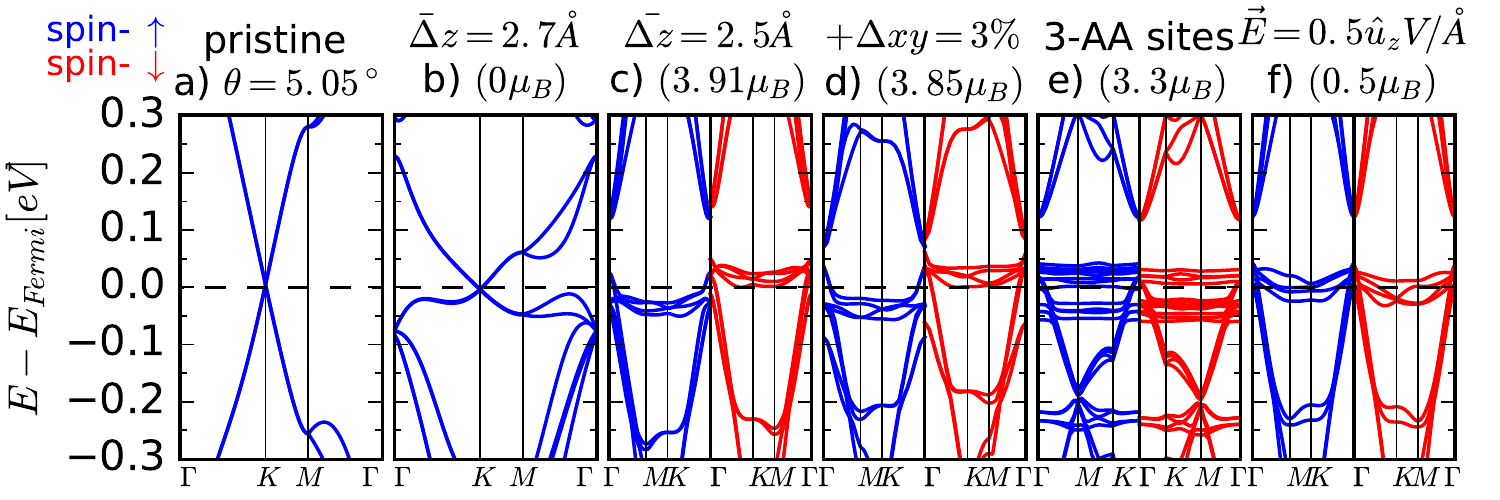}
 \caption{Electronic band diagrams of $i$=6 tBLG under different physical stimuli. The linear dispersion of the fully relaxed bilayer, a), is lost when interlayer distance $\bar{\Delta}z$ decreases in average to 2.7\AA, b). c) The system becomes magnetic for additional compression to 2.5\AA, exhibiting a magnetic moment of 3.97$\mu_B$ (Figure \ref{periodic}a). d) Biaxial strain modulates the magnetic moment to 3.85$\mu_B$. e) A three-AA-site supercell (Figure \ref{periodic}c) exhibits ferromagnetic alignment of the three 3.3$\mu_B$ magnetic moments. f) an external electric field applied perpendicularly to the bilayer plane balances the spin-polarized charge distribution creating magnetic moments locally anti-aligned (Figure \ref{periodic}d).}  
 \label{figBandsi6}
\end{figure*}

\section{\label{methods}Computational methods}
Self-consistent calculations of the electronic structure of pristine and doped tBLGs were performed with the localized atomic-like orbital basis set of the {\it SIESTA} density functional theory (DFT) based code \cite{PhysRevB.53.R10441,0953-8984-14-11-302}.
A double-$\zeta$ basis set within the local density approximation (LDA) approach for the exchange-correlation functional was used. Although vdW functionals are known to describe accurately the interaction between the parallel layers of pristine graphene sheets \cite{PhysRevLett.103.096}, LDA was considered a suitable functional since the dominating effects here are the new states coupling the graphene layers upon compression and chemical modifications. Also, within the LDA, the required level of accuracy in the description of large unit cells can be obtained at an affordable computational cost. Atomic positions of vdW-distance bonded bilayers were relaxed with a force tolerance of 1 meV/\AA. Integration over the Brillouin zone (BZ) of the smaller tBLGs was performed using a Monkhorst sampling of 18$\times$ 18 $\times$ 1 k points. The number of k-points used in larger cells were reduced accordingly to the length of the unit cell vector. The radial extension of the orbitals had a finite range with a kinetic energy cutoff of 50 meV to ensure a good overlap between parallel sheets. The numerical integrals were computed on a real space grid with an equivalent cutoff of 300 Ry.

\section{\label{results}Results and discussion}

Moir{\'e} lattices are created by choosing a twist angle between two graphene layers so that a unit cell with coincidence site lattice points is created\cite{Campanera,dosSantos,Shuyang}. Commensurate angles are parameterized by an integer $i$ where the twist angle $\theta_i$ is given by cos$\theta_i=\frac{3i^2+3i+0.5}{3i^2+3i+1}$ \cite{dosSantos,Stauber}. We start by considering four tBLG geometries in which one layer of AA-bilayer graphene is rotated an angle $\theta_i$ with respect to the other layer. Twist angles $\theta_{i=6}=5.09^\circ$, $\theta_{i=8}=3.89^\circ$, $\theta_{i=10}=3.15^\circ$, and $\theta_{i=16}=2.0^\circ$ were considered.
Internal coordinates of pristine and doped tBLGs separated by the vdW distance of 3.3\AA\ were fully relaxed. Pressure is simulated by reducing the separation between fully-relaxed parallel layers. Electronic and magnetic properties of compressed tBLG structures were computed with atoms at fixed positions, which is known to introduce minor changes upon additional relaxation\cite{Carr2018}. To avoid interactions between neighboring cells, tBLGs were modeled within supercells separated by 33.0\AA\ in the perpendicular-to-the-plane direction.

Figure \ref{figBandsi6}a shows the band structure of the fully relaxed $i$=6 tBLG. The two degenerate bands observed in the $\Gamma \rightarrow K \rightarrow M$ high-symmetry path of the BZ exhibit the linear dispersion of graphene and can be related to the absence of short-range correlation between layers, similarly to turbostratic bilayer graphene\cite{Latil2006,Latil2007}. After a rigid approach of one sheet to the other so that the vertical separation is reduced in average to $\bar{\Delta z}=2.7$\AA, the massless fermion character of the bilayer vanishes, as shown in Figure \ref{figBandsi6}b. This effect is a consequence of the enhanced hybridization between layers which induces localization, whereas band degeneracy is preserved due to the lack of symmetry breaking. This localization mechanism is a consequence of the potential introduced by the Moir{\'e} pattern and can be described by tight-binding calculations in low-twist angles\cite{PhysRevB.86.125413,Laissardiere}.

\begin{figure}[htp]
 \centering
   \includegraphics[width=0.48\textwidth]{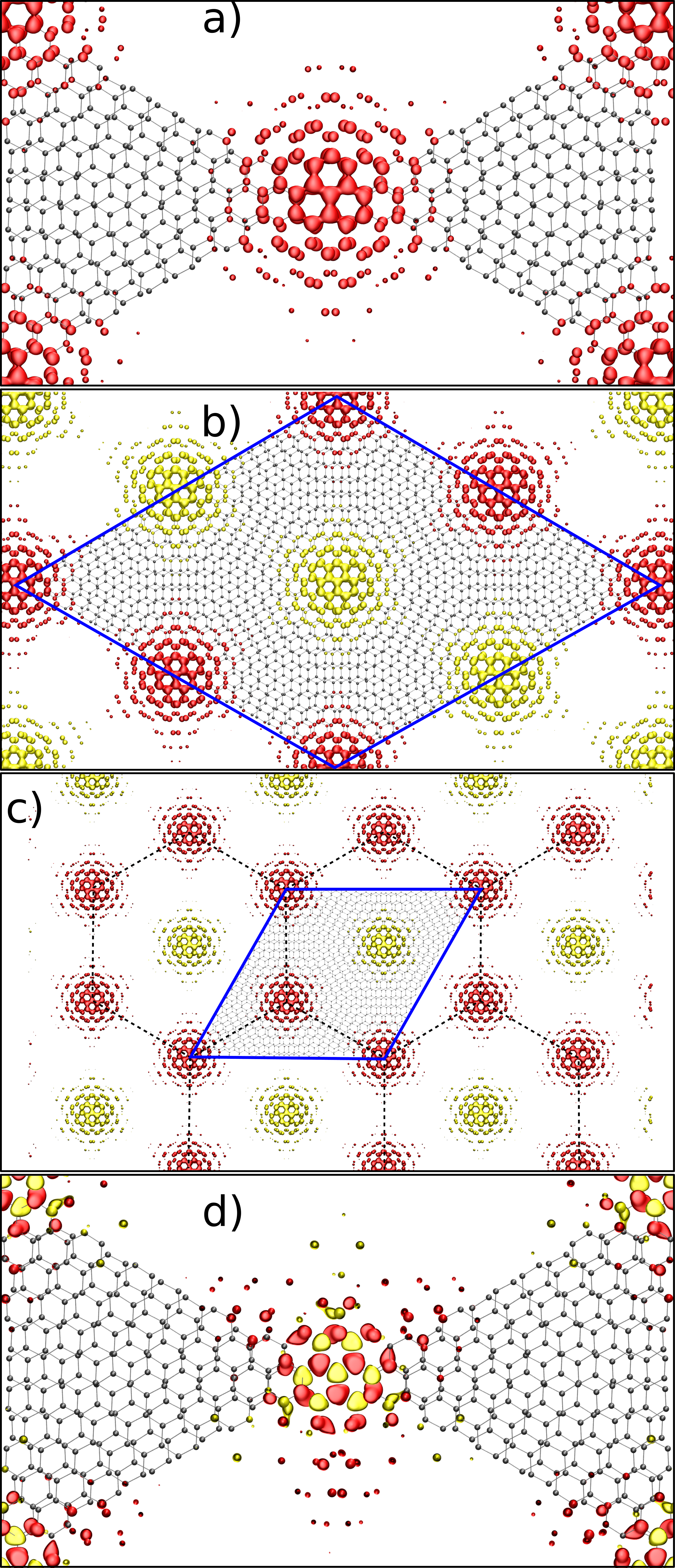}
   \caption{Real-space spatial distribution around the AA staking region of four spin configurations. a) Compressed $\theta_{i=6}=5.09^\circ$ tBLG creates a circular distribution of unbalanced charge. b) Extending four times the unit cell (blue-line polygon) ferromagnetic a single-Q magnetic state is observed. c) Ferrimagnetic ordering in a hexagonal supercell containing three-AA-sites (blue-line polygon). d) Application of an electric field perpendicular to the tBLG plane leads to a local anti-ferromagnetic ordering distributed circularly around the AA site. Isosurfaces correspond to a charge density of $\sim 10^{-3}$ e$^-$/\AA$^{3}$}  
 \label{periodic}
\end{figure}

Additional compression reducing the interlayer distance in average to $\bar{\Delta z}=$2.5\AA\ leads to a completely different scenario. The dispersive twofold-degenerate bands accommodating two electrons apiece collapse in two groups (one for each spin quantum number) of four nearly flat states (see Figure \ref{figBandsi6}c). Minority and majority spin flat-bands are completely empty and filled, respectively. A small fraction of charge that is held in the dispersive bands at lower energies prevents the bilayer from exhibiting half-metallicity. The tBLG becomes magnetic, and a total magnetic moment of 3.91 $\mu_B$ emerges centered on the AA stacking region, exhibiting a ferromagnetic ordering between neighboring atoms. The spatial extension of the magnetic moment is of $\sim$1 nm, as displayed in Figure \ref{periodic}a. The pressure at this interlayer distance is estimated in 2.12 GPa. Smaller distances between parallel virtual planes were observed to introduce substantial variations in the total energy and magnetic moments of the bilayers. A configuration with a magnetic moment of 3.7$\mu_B$ was obtained in a unit cell with a vertical separation of 25 \AA\ between repeated images. 

To observe a possible anti-ferromagnetic ordering across AA stacked regions, additional unit cells must be included in the simulation. Spin-polarized calculations on tBLGs containing four $i$=6 unit cells (2032 atoms) were conducted considering both restricted and unrestricted values of the total magnetic moment of the supercell. 
The result is shown in Fig\ref{periodic}b, where a network of magnetic moments exhibits stripes aligned ferromagnetically but with an anti-ferromagnetic ordering across $\pm$60$^\circ$ direction.
The structure factor of this configuration would give one Bragg peak out of the three $M$ points of the BZ. This so-called 'single-Q' state with alternating   magnetic oscillations in two directions was obtained by fixing the supercell magnetic moment to 0.
The formation energy of this configuration is 1 meV lower than a configuration with 16$\mu_B$, namely 4$\mu_B$ per AA site. Multiple other metastable configurations were obtained conducting restricted and unrestricted calculations, which yield spin configurations of 12 and 10.75 $\mu_B$, both with higher formation energies. 

Within a collinear model, a more conclusive analysis involves the inspection of a hexagonal lattice composed of three AA stacking regions. 
The resulting supercell (1524 atoms) exhibits a ferrimagnetic ordering where two AA regions are anti-ferromagnetically aligned with the third one, as displayed in Figure \ref{periodic}c. The total magnetic moment of this configuration is 3.77 $\mu_B$, and its formation energy is 9 meV lower than a configuration with all spins aligned ferromagnetically exhibiting half-metallicity. The total magnetic moment of the latter is 11.8 $\mu_B$, namely 3.9 $\mu_B$ per AA site, which recovers the result of Figure \ref{figBandsi6}c.
Considering that in the supercell containing three AA regions the magnetic moments are allowed to interact with more degrees of freedom, and due to the exchange-correlation energy favoring the ferrimagnetic alignment, one can conclude that the alignment of Figure \ref{periodic}c is preferred. 

Within the anti-ferromagnetic Heisenberg model (in which neighboring spins can antialign with any possible angle) tBLG is ripe for frustration owing to the triangular arrangement of neighboring magnetic moments. Not existing a physical way to maximally satisfy all three bonds in the triangular lattice simultaneously, a possible solution would entail a noncollinear spin alignment where spins would form a triangular formation at 120$^\circ$. This possibility is typical of materials hosting highly localized magnetic moments on $d$- or $f$-orbitals and less likely on delocalized C atom $p$-orbitals. Without an {\it ad hoc} on-site Coulomb repulsion term added in the DFT calculations, noncollinearity or spirals were not observed\cite{Arraga,Skyrmions}. 

A magnetic moment at an AA region can be slightly modulated by engineering the tBLG with biaxial strain, which induces a symmetric lattice deformation without breaking the system symmetry. Increasing the length of the hexagonal unit cell vector by 3\%, an enhanced chemical potential misalignment for both spin distributions is obtained. Figure \ref{figBandsi6}d shows that the two majority-spin bands that were partially filled become fully filled, whereas two of the flat states become partially empty. In the minority spin channel, one of the flat bands is slightly populated receiving the transferred charge. As a result, the tBLG reduces its net magnetic moment to 3.85 $\mu_B$.

Application of an electric field to bilayer AB graphene is a well-established method for opening an electronic band gap\cite{BLGefield}. Cao {\it et al.} \cite{Jarillo2} proposed the application of perpendicular electric fields by means of differential gating to tune interactions in magic-angle tBLG.
Here, it was observed that local ferromagnetic ordering at an AA site can be further tuned by applying a displacement field of 0.5 V/\AA\ perpendicularly to the tBLG plane. The total magnetic moment decreases to 0.5$\mu_B$ by equalizing the charge distribution across electronic states of both spin channels. Both the minority and the majority spins exhibit at the Fermi level (see Figure  \ref{figBandsi6}e) some of the valence bands partially filled and one fully empty, but with a misalignment in the chemical potential. Figure \ref{periodic}d shows the local magnetic moments extended anti-ferromagnetically at in-plane first-neighboring C atoms over the vicinity of the AA stacking region. The local moments align ferromagnetically with the closest atoms in the opposite layer. Increasing the electric field up to 1 V/\AA, or decreasing it down to almost zero slightly modifies the total magnetic moment. Removing completely the external field leaves the bilayer in the local anti-ferromagnetic configuration with a total magnetic moment of 0.4$\mu_B$, and a formation energy 20 meV above the ferromagnetic configuration discussed above (Figure \ref{periodic}a). Calculations were performed with flat graphene sheets, which was observed to lower the formation energy of the system upon compression plus electric field by several eV with respect to the fully relaxed and compressed structures.

\begin{figure}[htp]
 \centering
   \includegraphics[width=0.48 \textwidth]{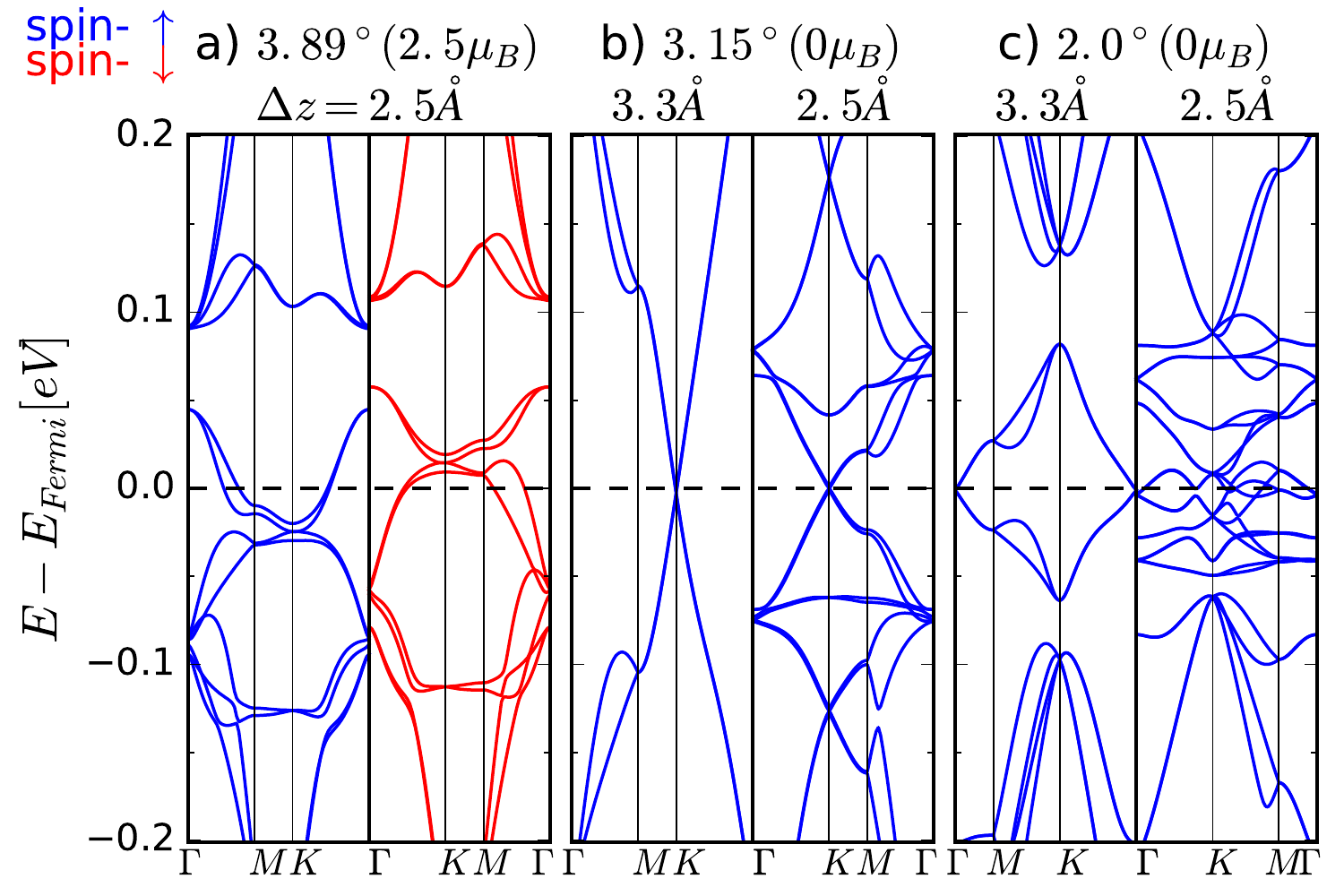}
 \caption{The spin-resolved band diagram of $\theta_{i=8}=3.89^\circ$ tBLG with graphene sheets separated by an average distance of 2.5\AA\ exhibits a total magnetic moment of 2.5$\mu_B$. Both b) $\theta_{i=10}=3.15^\circ$ and c) $\theta_{i=16}=2.0^\circ$ exhibit paramagnetic configurations. At 2.5\AA\ of interlayer separation, electronic states exhibit a much lower dispersion than at the relaxed distance of 3.3\AA. }  
 \label{figBandsi8}
\end{figure}

Decreasing the twist angle to $\theta_{i=8}=3.89^\circ$, the twofold degenerate bands of the paramagnetic vdW-bonded tBLG collapse at the Fermi level in four nearly flat bands of identical shape across the two spin channels when the layers are separated by an average distance of 2.5\AA.
An energy misalignment of some meV yields a net magnetic moment of 2.5$\mu_B$ (see Figure \ref{figBandsi8}a).  
Note the similarity of the bands at the Fermi level with those of ordinary AB stacked bilayer graphene \cite{Latil2006}, namely  only two bands touch each other at $K$ point with non-linear dispersion.
The spatial distribution of the magnetic moment is similar to the one shown for $i=6$ tBLG in Figure \ref{periodic}a.
Biaxial strain equivalent to increasing the unit cell vector in a 3\% modulates the electronic states of both charge distributions and reduces the magnetic moment to 2.0$\mu_B$. 

Interestingly, a slight decrease in the twist angle to $\theta_{i=10}=3.15^\circ$ in compressed tBLG leads to a paramagnetic spin distribution with no band misalignment.
Reducing the interlayer distance to $\bar{\Delta z}=2.5$\AA, the dispersion of the vdW-bonded bilayer bands is reduced from 0.5 eV to $\sim$0.15 eV (Figure \ref{figBandsi8}b). No disruption of the bands other than spanning a reduced energy range and some hybridization with flat bands in the zone center is observed. A linear crossing is observed at the $K$ point and a set of two bonding and antibonding nearly-flat states emerges at $\pm$60 meV. 

In agreement with previous reports based on tight-binding calculations\cite{Koshino}, the electronic spectrum of a smaller twisted-angle bilayer such as $\theta_{i=16}=2.0^\circ$ (3268 atoms) exhibits fourfold low-dispersive degenerate bands separated by an energy gap from the rest of the conduction and valence bands. This result is exclusive of a configuration where internal coordinates have been relaxed. 
Approaching rigidly one layer 0.8\AA\ to the other, localization effects are enhanced. As opposed to the tight-binding calculations performed in Ref. \cite{Carr2018}, this parameter-free DFT calculation demonstrates that no flat-band similar to the one observed at the magic angle is achieved upon reducing the interlayer separation. Pressure results in an reduction of the electronic states bandwidth and the accumulation of additional electronic bands in the vicinity of the Fermi level. No magnetic moment is created upon compression. Considering that additional calculations on $\theta_{i=5}=6.0^\circ$ showed that pressure causes the apparition of magnetism, it can be concluded that spontaneous magnetism in compressed tBLG arises for relatively large twist angles where $\theta_{i=10}=3.15^\circ$ is a lower limit. 

\begin{figure}[htp]
 \centering
   \includegraphics[width=0.48\textwidth]{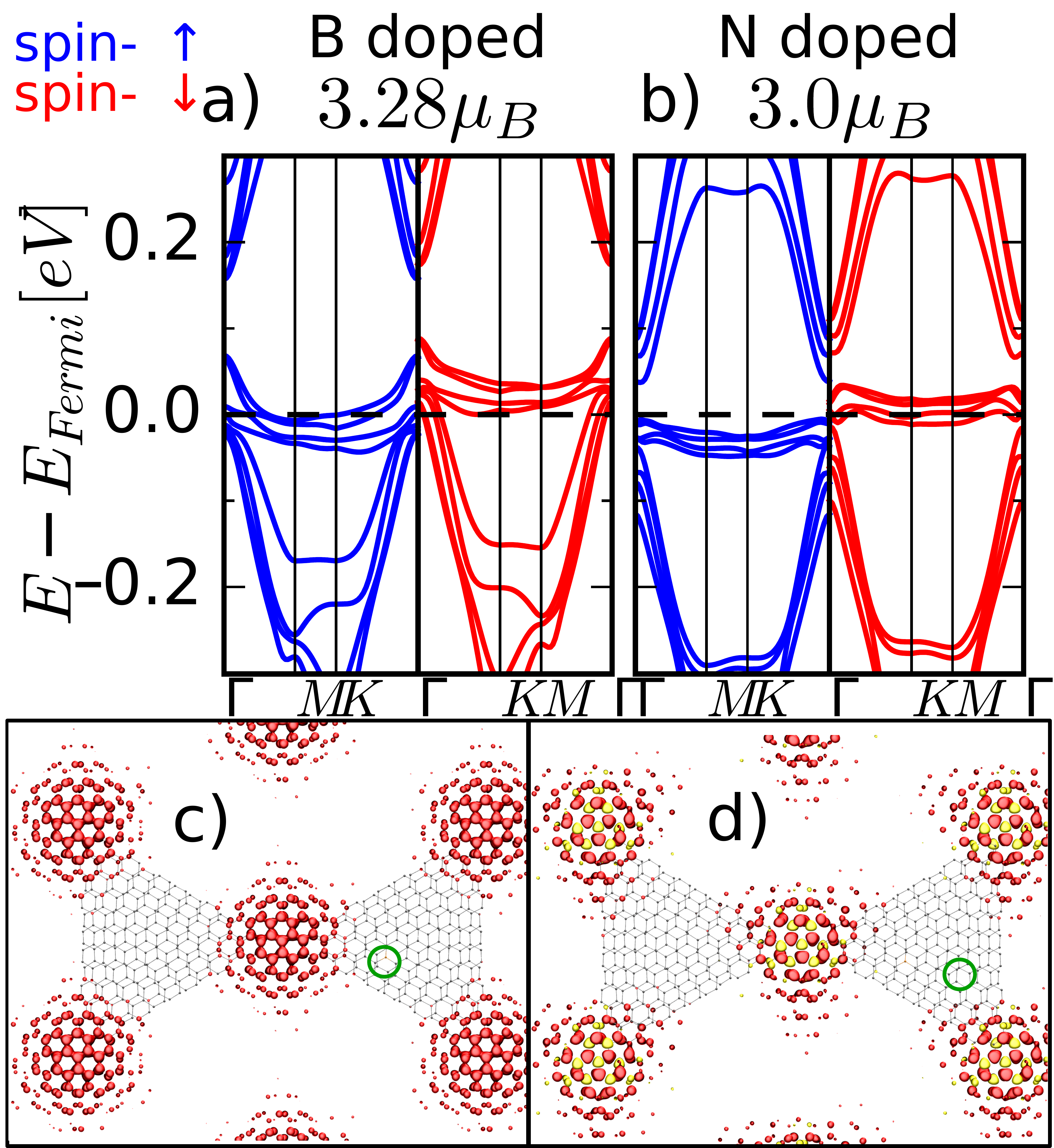}
 \caption{Spin-resolved band diagrams of $i$=6 tBLG doped with a) B, and b) N atoms. Half-metallicity is achieved with N doping, whereas B doping adds extra dispersion to the bands at the Fermi level. c) B doping leaves the original ferromagnetic alignment unchanged. d) N doping induces anti-ferromagnetic ordering between nearest neighboring atoms centered at the AA stacked region. Green circles point out to the doping atoms. Isosurfaces correspond to a charge density of $2*10^{-3}$ e$^-$/\AA$^{3}$}  
 \label{figBNdoping}
\end{figure}

Cao {\it et al.}. \cite{Jarillo2} proposed electrostatic gating for changing the charge carrier density and induce superconductor-metal transition in magic-angle tBLG. 
Substitutional B and N chemical doping is an efficient method to introduce hole and electrons in graphitic layers, at the expense of creating defect sites that may disrupt the local geometry of the sheets. 
Figure \ref{figBNdoping} shows the band diagrams of a $i$=6 tBLG with one B/N atom at a position far from the AA stacked C atoms. Geometries were fully relaxed prior to decreasing the interlayer distance. Across the two spin channels, both types of doping atoms generate an identical distribution of electronic states but with different energy alignment with respect to each other. B doping removes one electron out of the bilayer and decreases the total magnetic moment to 3.28$\mu_B$, adding some dispersion to the flat valence bands. Interestingly, despite the donor behavior of an N atom, the magnetic moment of the bilayer decreases to 3.0$\mu_B$, enhances the localized character of the valence bands, and turns the bilayer into a half-metal, with one spin channel metallic while the other spin channel features a 40 meV insulating band gap. 

The real-space representation of the unbalanced charge distribution show two different scenarios. Single N atom doping changes the local ferromagnetic ordering of the magnetic moment to anti-ferromagnetic. Figure \ref{figBNdoping}c shows that, even though the foreign atom sits far from the AA stacking region, the magnetic moments remain centered at the AA stacked C atoms. Triangular-shape magnetic moments have a spatial extension similar to the one of pristine tBLG in Fig\ref{periodic}a, and are practically insensitive to external electric fields.

Empirical-free DFT-based calculations of systems containing hundreds of atoms requires one to assume some approximations that might compromise the accuracy of the results. An insufficient number of k-points may introduce a bias in the Fermi level alignment and, therefore, in the exact value of the magnetic moment on spin-unrestricted calculations. Also, a reduced basis set in the {\it SIESTA} code calculation may describe poorly the C atoms orbital hybridization at the AA stacking region. Results presented above for pristine tBLG were reproduced with the more precise double-$\zeta$ polarized basis set and no significant differences were found, with the exception of the magnetic moment of compressed $i=6$ tBLG that was estimated in 2.8$\mu_B$. The all-spin-up three-AA site $i=6$ tBLG computed with the same basis set yields an average magnetic moment of 3.6$\mu_B$, close to the 3.9$\mu_B$ obtained above.

An analysis of the magnetic properties of compressed tBLG flakes was conducted to obtain a description of isolated magnetic moments, namely without possible coupling with neighboring moments in virtual cells, and to avoid a possible lack of accuracy derived from an insufficient sampling of the BZ. $\Gamma$-point calculations of isolated AA stacked regions were conducted on $\theta_{i=6}$ and $\theta_{i=10}$ tBLGs created by repeating two times the unit cell along each in-plane cell vector. Armchair edges were terminated with H atoms to remove chemically reactive dangling bonds. Only H atoms were relaxed in the pristine and doped bilayer configurations.

\begin{figure}[htp]
 \centering
   \includegraphics[width=0.48\textwidth]{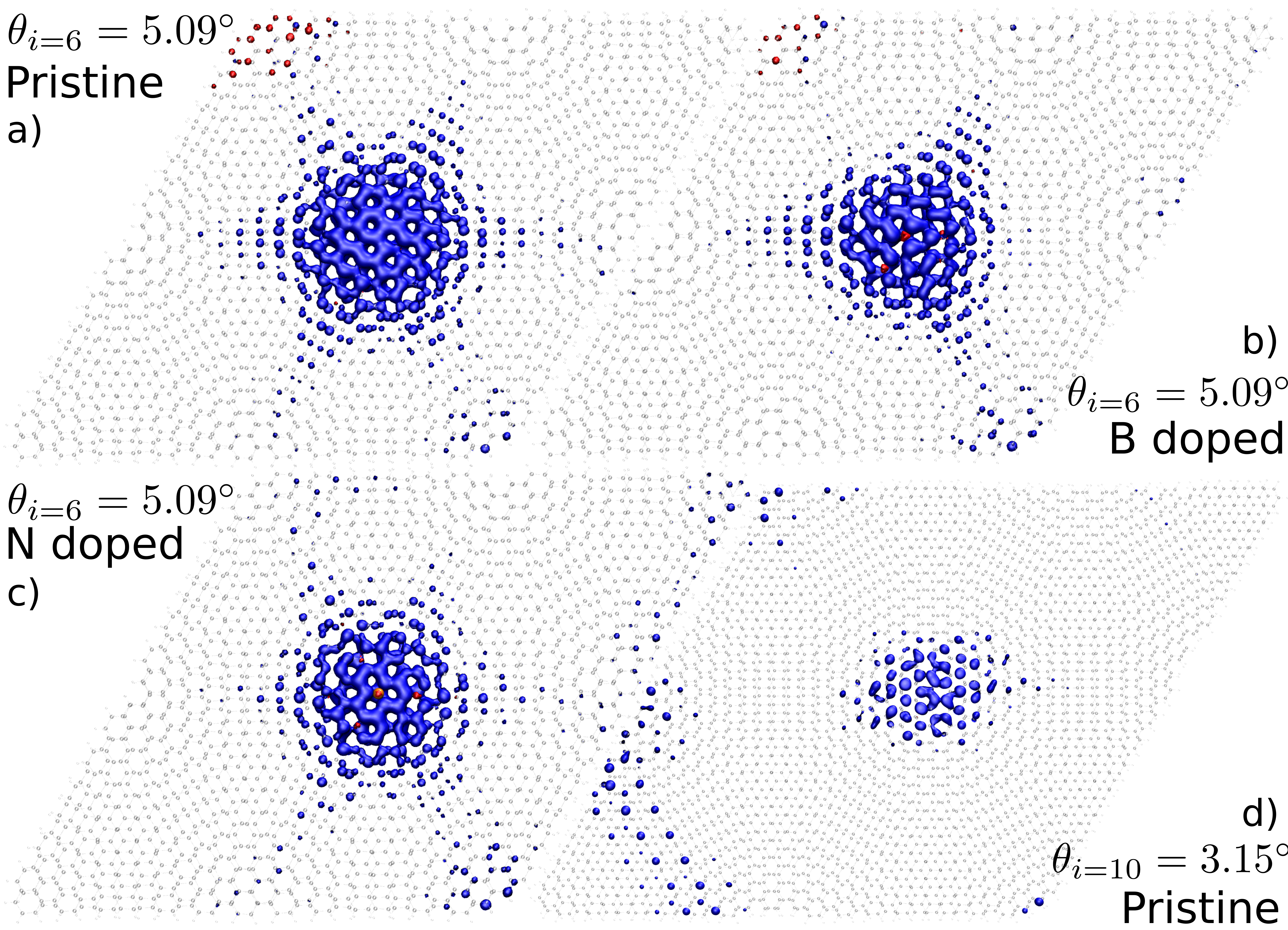}
 \caption{H atom-terminated flakes of compressed twisted bilayer graphene show that magnetism arises in non-periodic systems. Flakes are created by repeating four unit cells of $i$=6 and $i$=10 twisted bilayers. a) For a rotation angle of $\theta_{i=6}=$5.09 the spatial extension is larger than for a periodic system. Doping with B atoms b) change the shape of the magnetic moment to triangular. N doping c) reduces the spatial extension for the same isosurface of $10^{-3}$ e$^-$/\AA$^{3}$. d) shows the unbalanced charge distribution for $\theta_{i=10}=$3.15$^\circ$.}  
 \label{figSlabs}
\end{figure}

Spin-unrestricted calculations on a compressed $i$=6 tBLG flake yield a total magnetic moment of 3.0$\mu_B$ whose spatial extension increases with respect to the periodic arrangement to $\sim$1.5 nm. Figure \ref{figSlabs}a shows the circular arrangement of the magnetic moment in the vicinity of the central AA region with a six-vertex star shape ending as approaching the neighboring AA regions at the flake edges. Substitution of a C atom at the center of the flake with a B atom (one electron less than C) reduces the magnetic moment to 2.0$\mu_B$. The B atom and its 2$^{nd}$ and 4th neighboring C atoms are anti-ferromagnetically aligned to the rest of the network atoms, which contributes to changing the spatial extension of the magnetic moment to a three-vertex star shape (Fig\ref{figSlabs}b). A second B atom in the same layer at 4th neighboring position yields a total magnetic moment of 2.1$\mu_B$, whereas if it is in substitution of a C atom in the layer  below (14 meV lower in the formation energy), the moment is of 1.0$\mu_B$

One should expect that if doping with an acceptor atoms reduces the magnetic moment of the bilayer flake in 1$\mu_B$, doping with a donor N atom should have the opposite effect. However, substitution with a N atom at the AA site, and additional substitution with a second atom at 4th neighboring position, the same total magnetic moment as in B doping is obtained. A weak charge of opposite sign with ferromagnetic alignment between the N atom and its 4th nearest neighbors barely modifies the shape of the unbalanced charge distribution around the defect with respect to the pristine configuration (see Figure \ref{figSlabs}c). A compressed $i$=10 tBLG, that in the periodic configuration was paramagnetic, exhibits a 1$\mu_B$ magnetic moment. The AA region is isolated in a bilayer flake where edges were cut back so only one AA site is present. Figure \ref{figSlabs}d shows the distribution of the magnetic moment over a flake demonstrating that the absence of periodicity is not an impediment to the modulation of magnetism in tBLG.

\section{\label{Conclusions}Conclusions}
 
Combining uniaxial vertical pressure with a precise twist of the graphene sheets, dynamic band-structure engineering in graphene Moir{\'e} superlattices is possible, demonstrating that tBLG is a rich ground for hosting emergent flat-band magnetism and half-metallicity. The multiple degeneracies of the single-particle energy spectrum of tBLG that are due to inherent symmetries are broken when uniaxial vertical strain enhances electronic localization at AA stacking sites and induces rehybridization of local atomic orbitals creating magnetic ordered states.
A hexagonal lattice of AA sites has a preference for ferromagnetic ordering hosting an anti-ferromagnetic ordered magnetic moment in the center. In light of the small differences in energy observed, it would be adventurous to select one spin ordering over another to remove the observed practical degeneracy of the system's ground state. A critical conclusion leads us to note that, owing to accuracy limitations of DFT calculations with a large number of atoms, this result must be taken cautiously and the all-up spin configuration must not be totally discarded as the ground state of the system.

Even though the magnetic state of a tBLG can be sensitive to the accuracy of the simulation parameters (i.e., extension of the basis set, orbitals' energy cutoff, sampling of the BZ, etc), corrections to conducted DFT calculations, without particular assumptions about interactions among the charge carriers, is not expected to modify the ground state towards a frustrated anti-ferromagnetic order. Experimental observations are expected to shed light on the long-range ferromagnetic ordering of tBLG magnetic states.

The artificial injection of electrons into the sheets of experimental realizations was mimicked here in compressed tBLG by chemical doping with B and N atoms which modulates the magnetic moments and induces flat-band half-metallicity. An external electric field allows for local ferro- to anti-ferromagnetic ordering transition of spatially localized states of periodic systems, similar to N atom doping.
The vanishing bandwidth (flat states) of periodic magnetic arrangements was reproduced on finite-size bilayer flakes resulting on different magnetic moments depending on the twist angle. The emergence of magnetism on compressed tBLG is not related directly to the periodicity of the system but to the enhanced hybridization of the stacked C atoms.

\section{\label{acknowledgments}Acknowledgments}
Los Alamos National Laboratory is managed by Triad National Security, LLC, for the National Nuclear Security Administration of the U.S. Department of Energy under Contract No. 89233218CNA000001. This work was supported by the U.S. DOE Office of Basic Energy Sciences Program E3B5. I acknowledge the computing resources provided on Bebop, the high-performance computing clusters operated by the Laboratory Computing Resource Center at Argonne National Laboratory. 

\end{document}